\documentclass[reprint,superscriptaddress,amsmath,amssymb,aps,pra]{revtex4-2}
\usepackage[T1]{fontenc}
\usepackage{physics}
\usepackage[utf8]{inputenc}
\usepackage[english]{babel}
\usepackage{graphicx}     
\usepackage{hyperref}
\usepackage{xcolor}
\usepackage{soul}
\usepackage{float}
\usepackage[normalem]{ulem}

\hypersetup{colorlinks=true,citecolor={blue},linkcolor={blue},urlcolor={blue}}

\begin{document}

\title{Phase-Space  Framework for Noisy Intermediate-Scale Quantum  Optical Neural Networks}

\author{S.~Świerczewski}
\affiliation{Institute of Experimental Physics, Faculty of Physics,\\University of Warsaw, ul. Pasteura 5, PL-02-093 Warsaw, Poland}
\author{W. Verstraelen}
\affiliation{Division of Physics and Applied Physics, School of Physical and Mathematical Sciences, Nanyang Technological University, Singapore, Singapore}
\affiliation{Majulab, International Joint Research Unit UMI 3654, CNRS, Université Côte d’Azur, Sorbonne Université, National University of Singapore, Nanyang Technological University, Singapore, Singapore 117543}
\author{P. Deuar}
\affiliation{Institute of Physics, Polish Academy of Sciences,\\ Aleja Lotnik\'ow 32/46, PL-02-668 Warsaw, Poland}
\author{B.~Pi\k{e}tka}
\affiliation{Institute of Experimental Physics, Faculty of Physics,\\University of Warsaw, ul. Pasteura 5, PL-02-093 Warsaw, Poland}
\author{T. C. H. Liew}
\affiliation{Division of Physics and Applied Physics, School of Physical and Mathematical Sciences, Nanyang Technological University, Singapore, Singapore}
\affiliation{Majulab, International Joint Research Unit UMI 3654, CNRS, Université Côte d’Azur, Sorbonne Université, National University of Singapore, Nanyang Technological University, Singapore, Singapore 117543}
\author{M.~Matuszewski}
\affiliation{Center for Theoretical Physics, Polish Academy of Sciences,\\ Aleja Lotnik\'ow 32/46, PL-02-668 Warsaw, Poland}
\affiliation{Institute of Physics, Polish Academy of Sciences,\\ Aleja Lotnik\'ow 32/46, PL-02-668 Warsaw, Poland}
\author{A.~Opala}
\email{opala@fuw.edu.pl}
\affiliation{Institute of Experimental Physics, Faculty of Physics,\\University of Warsaw, ul. Pasteura 5, PL-02-093 Warsaw, Poland}
\affiliation{Institute of Physics, Polish Academy of Sciences,\\ Aleja Lotnik\'ow 32/46, PL-02-668 Warsaw, Poland}

\begin{abstract}
Quantum optical neural networks (QONNs) enable information processing beyond classical limits by exploiting the advantages of classical and quantum optics. However, simulation of large-scale bosonic lattices remains a significant challenge due to the exponential growth of the Hilbert space required to describe a quantum network accurately. Consequently, previous theoretical studies have been limited to small-scale systems, leaving the behaviour of multimode QONNs largely unexplored. This work presents an efficient computational framework based on the phase-space positive-$\mathcal{P}$ method for simulating bosonic neuromorphic systems. This approach provides a view to previously inaccessible regimes, allowing the validation of large-scale bosonic networks in various quantum machine learning tasks such as quantum state classification and quantum state feature prediction. Our results show that the performance of a large quantum reservoir does not improve monotonously with the number of bosonic modes, instead following a complex dependence driven by the interplay of nonlinearity, reservoir size, and the average occupation of the input mode. These findings are essential for designing and optimising optical bosonic reservoirs for future quantum neuromorphic computing devices.
\end{abstract}

\maketitle

\section{Introduction}
\textit{Optical neural networks} (ONNs) have emerged as a key platform for neuromorphic computing in both quantum and classical domains. ONNs enable ultra-fast (THz-range) and energy-efficient (below 10 fJ) information processing~\cite{Ma_2025, Matuszewski_2024}. At the same time, ONNs are inherently sensitive to quantum signatures encoded in optical fields ~\cite{Roberta_2024}. These unique properties offer promising prospects for implementing \textit{ quantum neuromorphic computing} (QNC) with optical devices.

QNC arises from the synergy between brain-inspired information processing and quantum hardware~\cite{Ghosh_2019, Marković_2020, Ghosh_2021}. By leveraging quantum data interpretation, optical QNC opens up new frontiers for next-generation computing and sensing devices~\cite{Roberta_2021, Yao_2025}. However, training and scaling quantum optical neural networks remains an open scientific challenge. Therefore, in recent years, a promising solution based on \textit{quantum reservoir neural networks} (QRNNs), merging classical reservoir computing paradigms with quantum systems, has been extensively explored \cite{Ghosh_2021}.

Unlike classical neural networks, QRNNs can process quantum correlations inherent in an input state's density matrix $ \hat{\rho} $. These correlations, quantified by measures such as concurrence $ C(\hat{\rho}) $, entanglement entropy $ S(\hat{\rho}) $, or negativity $ \mathcal{N}(\hat{\rho}) $, are indirectly embedded in the reservoir dynamics \cite{Ghosh_2019}. Interestingly, a simple linear classifier, forming an output layer of a quantum reservoir neural network, can learn to extract quantum properties by analysing the average occupancy of reservoir nodes. Thus far, QRNNs have been theoretically explored for quantum state classification~\cite{Ghosh_2019}, feature prediction \cite{Krisnanda_2022,ko_2025}, state generation \cite{Ghosh_2019_prep, Krisnanda_2021, Xu_2023}, tomography \cite{Krisnanda_2023TM}, entanglement sensing~\cite{Krisnanda_2023, Krisnanda_2023WC}, or quantum circuit implementation~\cite{Ghosh_2021QC}.

The general theoretical framework used to describe quantum reservoirs is typically based on the density matrix formalism. In this approach, the Hilbert space of the reservoir is constructed as a tensor product of $N$ local Hilbert spaces $ \mathcal{H}_d$  (or truncated Fock spaces), representing each reservoir site. For fermionic sites, $\mathcal{H}_d$ is inherently limited to two-dimensional spaces $d = 2 $ due to the Pauli exclusion principle, enforcing a binary occupancy at each lattice node. In contrast, bosonic sites require flexible truncation levels $d$, reflecting the maximum particle occupation allowed per reservoir mode. Consequently, the computational Hilbert space scales as $\text{Dim}(\mathcal{H}) \sim d^{2N}$, including all mixed states. Therefore, capturing higher-order quantum correlations encoded in bosonic systems, requiring high truncation levels, leads to significant computational challenges. In practice, even for few-mode reservoirs, the modelling of the system can be practically impossible if the nodes are highly populated, or if the correct modelling of a quantum state requires high Fock space truncation (for example, for the Schrödinger cat quantum states~\cite{Kirchmair_2013, Grimm_2020}).

The large size of the computational Hilbert space poses a significant challenge, particularly during the simulation of reservoirs trained using supervised learning, as it requires processing many samples to ensure generalisation and mitigate overfitting~\cite{dawid_2023}. These demands create a computational bottleneck, particularly during the training of multi-mode QNC systems, with input represented as quantum states. Therefore, previous works have considered quantum reservoirs containing only a few bosonic \cite{Xu_2023} or only 
fermionic sites \cite{Ghosh_2019}. These constraints hinder the theoretical study of large-scale bosonic quantum reservoirs and pose challenges for the theoretical validation of QNC devices.\\

In this work, we explore the \textit{Positive-$\mathcal{P}$ Method} (PPM), a stochastic phase-space approach, as a computational framework that enables the simulation of ``large-scale'' bosonic quantum systems (more than dozens of nodes). Quantum networks of this scale are commonly classified as noisy intermediate-scale quantum systems~\cite{Preskill_2018,Cheng_2023}. By leveraging PPM, we demonstrate the performance of large-scale photonic reservoirs in tasks such as quantum state classification (QSC) and quantum future prediction (QFP). We identify a key limitation of large quantum reservoir performance, which arises when the input quantum information is dispersed across too many reservoir nodes. This effect reduces the effective nonlinearity of the reservoir, leading to the degradation of computational efficiency. 
Performing reservoir optimisation, we find an \textit{optimal balance} that harmonises reservoir size and input field density in a given quantum task.

\section{Model} \label{sec:model}
\begin{figure*}[bht!]
    \centering
    \includegraphics[width=1\linewidth]{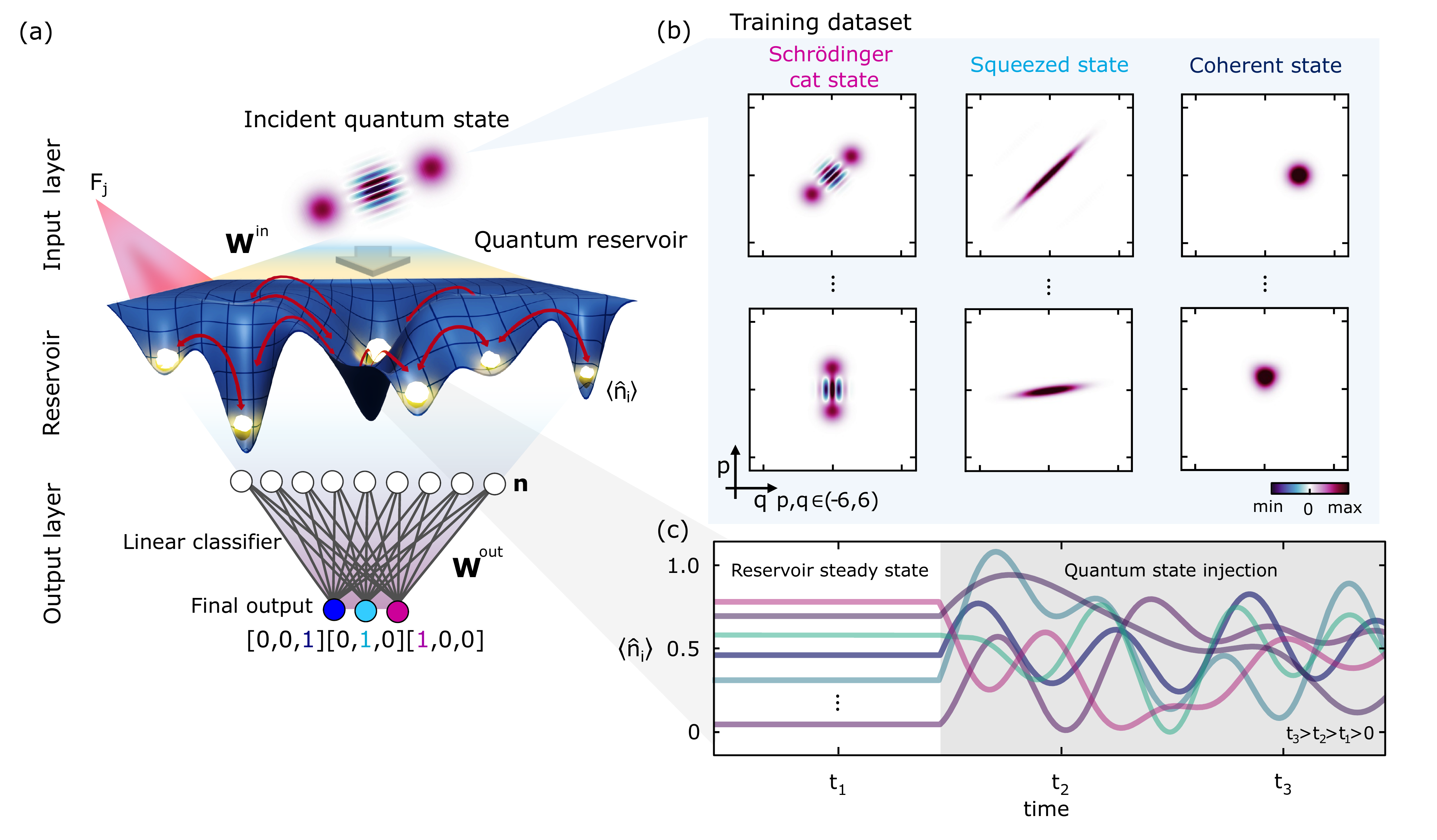}
    \caption{\textbf{Schematic illustration of a quantum reservoir applied for quantum state recognition.} (a), Structure of the QRNN, depicting input, reservoir, and output layers. The reservoir contains nine nonlinear bosonic nodes that transform input quantum states into real-valued observables given by average occupations $\langle \hat{n}_i \rangle$, where $i$ indexes the nodes. (b), samples from the training dataset consisting of various quantum states, represented as Wigner functions where $\mathcal{W}(q,p)$, $q$ and $p$ denote position and momentum quadratures, respectively~\cite{Wigner_1932}. Input quantum states were categorised into three classes (Schrödinger cat states, squeezed states and coherent states), labelled in the output layer using a \textit{one-hot encoding}. The output weights $\mathbf{W}^{\text{out}}$, defining the connections between the reservoir and the output neural network layer, are optimised to classify the states in a supervised learning process. (c), Schematic dynamics of reservoir states, when the incident quantum state perturbs the reservoir’s steady state, maintained by continuous coherent driving fields $F_j$.}
    \label{fig:fig1}
\end{figure*}

\subsection{Quantum reservoir computing} \label{sec:modelA}
Figure~\ref{fig:fig1} illustrates a QRNN, constructed as a set of coupled bosonic modes, designed for quantum state classification. During QSC, input information, encoded in \textit{quantum light}, is injected into the reservoir. This perturbation induces a dynamical response in the bosonic modes, effectively processing the input quantum state within the high-dimensional Hilbert space of the reservoir modes, as schematically shown in Fig.~\ref{fig:fig1}c. In such a case,  the complete Hilbert space of the system can be defined as
\begin{equation}
\label{eq:1}
    \mathcal{H} =  \mathcal{H}_{d'}^\text{in} \otimes\mathcal{H}_d \otimes\mathcal{H}_d \otimes   \dots \otimes  \mathcal{H}_d = \mathcal{H}_{d'}^\text{in} \otimes\bigg(\bigotimes_{i=1}^{N}  \mathcal{H}_d\bigg),
\end{equation}
where $\mathcal{H}_{d'}^\text{in}$ represents the input state Hilbert space, defined with truncation $d'$.
Analogous to classical machine learning, this process can be interpreted as a \textit{feature space expansion}.

To fully harness the benefits of quantum reservoir computing (QRC) over linear information processing (possible when a quantum processor is created via linear components), the Hamiltonian $ \hat{H}_{R} $ should incorporate higher-order terms. Such nonlinearities can arise due to presence of \textit{Kerr-like} nonlinearity $\frac{U}{2}\hat{b}^\dagger \hat{b}^\dagger \hat{b} \hat{b} $, \textit{cross-Kerr} interactions $ U_{(1,2)} \hat{b}_1^\dagger \hat{b}_1 \hat{b}_2^\dagger \hat{b}_2 $~\cite{Kounalakis_2018}, or \textit{parametric processes $ \alpha \chi (\hat{b}^2 + \hat{b}^\dagger{}^2) $}, where $b_{j}$ and $b_{j}^{\dagger}$ are respectively the annihilation and creation operators defined for reservoir site $j$.

These effects are intrinsically linked to the specific hardware realisation of the quantum reservoir. The presence of nonlinearity introduces complex behaviour within the reservoir, which can induce intensity-dependent phase shifts, cross-mode coupling, and spontaneous generation of quantum squeezing or even entanglement ~\cite{Cramer_2013}.

The result of nonlinear dynamics can be quantified by measuring average occupations of reservoir modes, expressed as $\langle \hat{n}_i \rangle = \text{Tr}(\hat{\rho} \hat{b}_i^\dagger \hat{b}_i).$
This measurement provides indirect access to the correlations encoded in the input state density matrix, facilitating the extraction of quantum features from the input states.  The measured average occupation at each node (at given time $t$), represented as vector $\mathbf{n}$, is then employed as a feature vector to classify quantum states through supervised learning. The reservoir density can be measured at different times to increase the size of the reservoir output (state) vector $\mathbf{n} = [\langle \hat{n}_1(t_1) \rangle, \langle \hat{n}_2(t_1) \rangle, \cdots, \langle \hat{n}_N(t_n) \rangle]$. This procedure,  called \textit{time multiplexing}, helps capture the increase in quantum correlation during the evolution of the reservoir. This allows quantum processing using small reservoirs to mitigate the computational difficulties of simulating large multimode systems.  However, performing time multiplexing requires high temporal measurement resolution, which is experimentally challenging. In this work, we focus on the more realistic realisation of QRC without time multiplexing, meaning that the size of the vector of features is the same as that of the studied reservoir. Instead of selecting specific times in the dynamics, we calculate the difference between the time-integrated emission from each reservoir mode and its steady-state value, $\langle \hat{n}_{i}^{s}\rangle$. We construct the vector $\mathbf{n}$ where $n_i = \int_{\Delta t} \langle \hat{n}_i(t) \rangle dt - \langle \hat{n}_{i}^{s}\rangle\Delta t$, and $i$ runs over all reservoir nodes.

Following the \textit{reservoir computing} (RC) paradigm, the proposed neuromorphic system consists of an input layer, a reservoir layer (hidden layer), and an output layer~\cite{Tanaka_2019}. Unlike traditional feedforward neural networks, in the RC framework, only the output layer is trained by optimising the output weight matrix $\mathbf{W}^{\text{out}}$. The input weights matrix $\mathbf{W}^{\text{in}}$, which describes the coupling between the source and the reservoir, and the linear coupling matrix $\mathbf{J}$, which defines the magnitude of mutual interactions between the reservoir nodes, remain constant during training. The network output is given by $\mathbf{Y} = \mathbf{W}^{\text{out}}\mathbf{n}$. The output is further employed using simple linear methods to perform a specific task. A detailed description of the learning process used for QSC and QFP is provided in the Appendix~\ref{sec:appA}.

\subsection{QRC with photonic systems}
The non-equilibrium nature of photonic neural networks demands a description considering the finite lifetime of photons. This can be done, for example, by employing the Lindblad master equation or phase-space framework. Therefore, the following sections introduce more precise formalisms for dissipative bosonic neuromorphic systems, including both frameworks.

In the absence of thermal excitations, the Lindblad master equation describing the  evolution of the reservoir density matrix $\hat{\rho}$ takes the following form 
\begin{equation}
    \label{Eq:1}
\dot{\hat{\rho}}\left(t\right)=-\mathrm{i}\left[{\hat{H}},\hat{\rho}(t)\right]+\sum_{j=1}^N\frac{\gamma_j}{2}\mathcal{L}(\hat{b}_j),
    \end{equation}
where $\hat{b}_j$ is the bosonic annihilation operator for reservoir site $j$. For simplicity, we begin with describing the dynamics of the studied quantum reservoir, however a more detailed model incorporating irreversible interactions between the reservoir and external source nodes via \textit{cascade coupling}~\cite{Lopez_2016, Ghosh_2019}, supported by numerical simulations, will be introduced in the section~\ref {QRCstates}.

In equation~(\ref{Eq:1}) $\mathcal{L}(\cdot)$ denotes a Lindblad superoperator,
which for a given field operator $\hat{b}$ acts on a density matrix as $\mathcal{L}(\hat{b})=2\hat{b}\hat{\rho}\hat{b}^\dag-\hat{\rho}\hat{b}^\dag \hat{b}-\hat{b}^\dag \hat{b}\hat{\rho}$.

For simplicity, we used units where $\hbar=1$.
The first RHS term of Eq. (\ref{Eq:1}) defines the conservative evolution given by the total Hamiltonian $\hat{H}$ (including input and reservoir nodes). In contrast, the second term introduces dissipation through the interaction with the external environment, occurring for each site with rate $\gamma_j$.

We consider a reservoir as a quantum Bose-Hubbard system arranged in a 2D square lattice,  with the Hamiltonian defined as
 \begin{equation}
 {\hat{H}}=\sum_{j}{\hat{H}}_j-\sum_{\langle ij \rangle}J_{ij}({{\hat{b}}_i}^\dag{{\hat{b}}_j}+{{\hat{b}}_j}^\dag{\hat{b}}_i),
 \label{Eq:2}
\end{equation} 
where $H_j$ is the local part of the Hamiltonian at site $j$, $\langle ij \rangle$ denotes the connections between the nodes, $J_{ij} = J_{ji}$ is a random nearest-neighbour Hermitian hopping between reservoir nodes. The summation was performed over all sites $i$ connected to $j$. We assume that the reservoir is driven by a coherent field $F_je^{-i\omega t}$, with complex amplitude $F_j$ and frequency $\omega$.
In the reference frame rotating with the laser frequency, the local part of the Hamiltonian with Kerr-type nonlinearity is given by
\begin{equation}
 {\hat{H}}_j=-\Delta_j{{\hat{b}}_j}^\dag{\hat{b}}_j+\frac{U}{2}{\hat{b}}_j^\dag{\hat{b}}_j^\dag{\hat{b}}_j{\hat{b}}_j+\sum_j\left(F_j \hat{b}_j^{\dagger}+F_j^* \hat{b}_j\right),
 \label{eq:33}
\end{equation} 
where $\Delta_j=\omega -\omega_j$, is the local energy detuning, given by the difference between resonant excitation source frequency and natural (randomly distributed)  reservoir mode frequency 
$\omega_j$. 

The parameter $U$ defines the strength of the local two-body interaction.

{\subsection{Positive-$\mathcal{P}$ method}}
Numerical integration of equation (\ref{Eq:1}) for the large-scale bosonic reservoir is computationally inefficient. Therefore, to solve this issue, we apply a \textit{phase-space}  approach employing the positive-$\mathcal{P}$ method~\cite{Drummond_1980, Deuar_2002, Deuar_2006,Gardiner_Zoller_2010}. The PPM has been successfully utilised to describe quantum dynamics in various physical platforms, including quantum optics ~\cite{Carter87}, 
Bose-Einstein condensates ~\cite{Drummond99, Carusotto01a, Deuar07,Perrin_2008,Kheruntsyan12,Lewis-Swan14},
Schwinger bosons~\cite{Ng_2011}, and exciton-polaritons, dealing with up to  tens and hundreds of thousands of nodes ~\cite{Deuar_2021,Deuar07}.

Phase space methods aim to represent the density matrix similar to a probability function over classical phase space. The Positive $\mathcal{P}$ function can describe a true probability density function (unlike e.g. the Wigner quasiprobability), but over a doubled phase space $(\alpha,\tilde{\alpha}^*)$. As such, its evolution is governed by a Fokker-Planck equation. This is known (by the Feynman-Kac theorem) to be equivalently recast as stochastic sampling over individual trajectories $(\alpha,\tilde{\alpha}^*)$. Notably, these trajectories can be independently evolved at minimal memory cost.

In this method, the necessity of describing the entire density matrix to calculate the quantum expectation values is then replaced by the calculation of averages over $\mathcal{S}$ stochastic trajectories in phase-space. In the limit of large trajectory number ($\mathcal{S}  \rightarrow \infty$), numerical calculations of the observable become asymptotically exact.

In the PPM, the density matrix $\hat{\rho}$ is more precisely expanded in terms of coherent state kernels, where phase-space variables and coherent states $\ket{\alpha}$ follow the direct dependence $\ket{\alpha}=\exp{\alpha \hat{b}^\dag-\alpha^*\hat{b}}\ket{0}$, where $\ket{0}$ is the vacuum state and $\hat{b}$ is the bosonic annihilation operator~\cite{Drummond_1980}.
The density matrix for the $N$-site 
bosonic reservoir is written as
\begin{equation}
    \hat{\rho}=\int{d^{2N}}{\boldsymbol{\alpha}}\,{d^{2N}}\tilde{{\boldsymbol{\alpha}}}\,\mathcal{P}({\boldsymbol{\alpha}},{\boldsymbol{\tilde{\alpha}}}^*)\,\hat{\Lambda}({\boldsymbol{\alpha}},{\boldsymbol{\tilde{\alpha}}}^*),
    \label{eq:dminP}
\end{equation}
where vectors $\boldsymbol{\alpha}$, $\boldsymbol{\tilde{\alpha}}^*$ contain all local phase-space variables $\alpha_j$ and $\tilde{\alpha}^*_j$ respectively,  $\hat{\Lambda}$ is the global projection operator, defined as 
$\hat{\Lambda}=\bigotimes_{i=1}^{N}\hat{\Lambda}_j(\alpha,\tilde{\alpha}^*)$
with unit trace, and $\hat{\Lambda}_j$ is the local coherent state projector kernel
$\hat{\Lambda}_j=\ket{\alpha_j}_j\bra{\tilde{\alpha}_j}_j/\braket{\tilde{\alpha}_j}{\alpha_j}$~\cite{Deuar_2021}. The function $\mathcal{P}({\boldsymbol{\alpha}},{\boldsymbol{\tilde{\alpha}}}^*)$ defines a real, positive probability distribution of stochastic configurations $\vec{v}=\{\boldsymbol{\alpha},{\boldsymbol{\tilde{\alpha}}}^*\}$.
With this representation, Eq.~(\ref{Eq:1}) can be directly mapped to a stochastic equation of motion \cite{Deuar_2021}. For a driven dissipative bosonic lattice within the tight-binding approximation, these equations take the form
\begin{align}
\label{eq:PP}
\begin{split}
    \left.\frac{\partial \alpha_j}{\partial t}\right|_{R} = \mathrm{i}\Delta_j \alpha_j - \mathrm{i}U \alpha_j^2\tilde{\alpha}_j^* - \mathrm{i}F_j - \frac{\gamma_j}{2}\alpha_j  \\
     + \sqrt{-\mathrm{i}U}\alpha_j\xi_j(t) + \mathrm{i}\sum_k J_{kj}\alpha_k,\\
    \left.\frac{\partial \tilde{\alpha}_j}{\partial t}\right|_{R} = \mathrm{i}\Delta_j \tilde{\alpha}_j - \mathrm{i}U \tilde{\alpha}_j^2\alpha_j^* - \mathrm{i}F_j - \frac{\gamma_j}{2}\tilde{\alpha}_j \\
     + \sqrt{-\mathrm{i}U}\tilde{\alpha}_j\tilde{\xi}_j(t) + \mathrm{i}\sum_k J_{kj}\tilde{\alpha}_k,
\end{split}
\end{align}
where $k$ sums over all connected sites. The $\xi_j(t)$ and $\tilde{\xi}_j(t)$ are independent real white noise (Wiener processes), fulfilling the following relations $\langle\xi_j(t)\xi_k(t)\rangle_S$=0, $\langle\xi_j(t)\xi_k(t')\rangle_S=\delta(t-t')_{jk}$ and $\langle\tilde{\xi}_j(t)\tilde{\xi}_k(t')\rangle_S=\delta(t-t')_{jk}$, where $\delta(\cdot)$ is a Dirac delta function and
$S$ denotes the system configurations taken into the stochastic average $\langle \cdot\rangle_S$ \cite{Deuar_2021}.
\newline{}
\begin{figure*}[t!]
\includegraphics[width=0.99\linewidth]{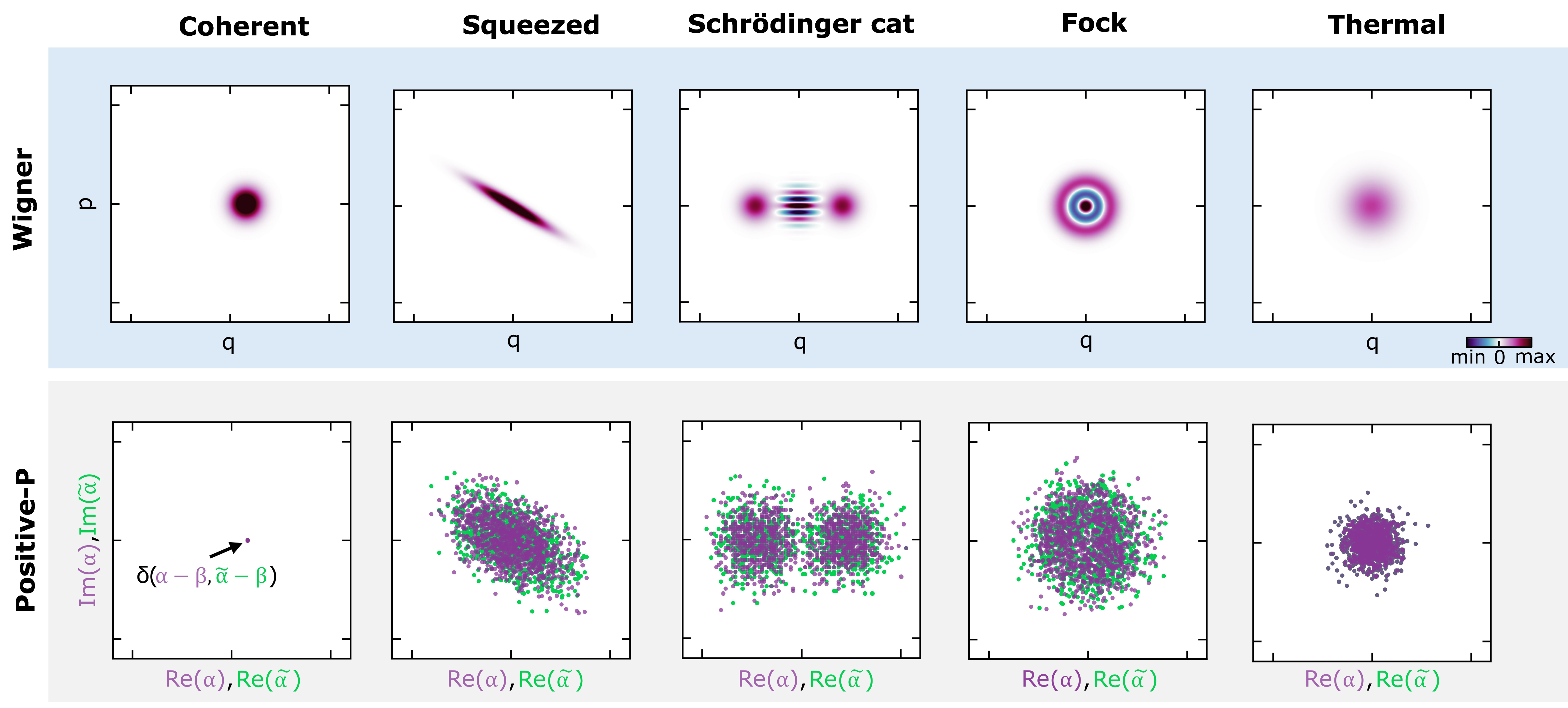}
    \caption{\textbf{Comparison of Wigner $\mathcal{W}(q,p)$ (top row) and positive-$\mathcal{P}\left(\alpha,\tilde{\alpha}^{*}\right)$ distributions (bottom row) for selected quantum states}. To visualise the positive-$\mathcal{P}$ distributions, the complex parameters $\alpha$ and $\tilde{\alpha}$ are separated into their real and imaginary parts, forming two-dimensional projections. Blue and purple points represent positions of $\tilde{\alpha}$ and $\alpha$, respectively (in reduced phase space). The number of stochastic trajectories used to represent each distribution varies depending on the state, from a single trajectory (coherent state) to several thousand trajectories (Schrödinger cat state). The coherent and thermal states have $\alpha=\tilde{\alpha}$, hence only one colour is visible.}
    \label{fig:fig8}
\end{figure*}

\subsection{QRC under the phase-space approach} \label{QRCstates}
In this section, we briefly describe the three main 
points required to implement reservoir processing in PPM formalism numerically: (I) \textit{Quantum State Sampling}, i.e., generating initial states according to correct phase-space distributions; (II) \textit{Input Signal Injection}, i.e., coupling input states to the quantum reservoir; and (III) \textit{Output Interpretation}, i.e., translating evolved quantum states into measurable outcomes by averaging over stochastic trajectories.

\textit{I. Quantum State Sampling}. Within the phase-space formalism, instead of considering the density matrix $\hat{\rho}$ of the input state, we describe the quantum state in terms of a probability distribution over the complex $(\alpha,\tilde{\alpha}^*)$ space. As an illustrative example, we introduce the sampling of coherent states (\textit{trivial case}). 

A coherent state $\ket{\beta}$, defined by the eigenvalue equation $\hat{b} \ket{\beta} = \beta \ket{\beta}$, represents the simplest sampling scenario. The phase-space representation of such states follows a well-known pseudo-probability distribution
\begin{equation}
 P(\alpha,\tilde{\alpha}^*) = \delta(\alpha - \beta) \delta(\tilde{\alpha} - \beta
 ).   
\end{equation}
In the equation above, a single pair of complex conjugate variables $\tilde{\alpha}=\alpha = \beta$ represents a quantum state.

As an example of more sophisticated (\textit{non-trivial}) quantum states, requiring stochastic sampling, we consider squeezed vacuum and optical Schrödinger cat states. Accordingly, it should be noted that the positive-$\mathcal{P}$ distribution is not unique due to multiple mathematically valid decompositions of the quantum state into a positive probability distribution in phase-space, satisfying the required representation of the density matrix.
However, it is possible to express a \textit{canonical} form of the positive-$\mathcal{P} $ distribution \cite{PhysRevA.43.1153}, which is directly connected to the Husimi-$\mathcal{Q}$ distribution, known to be real and positive. 

The positive-$\mathcal{P}$ distribution for a squeezed vacuum state, defined as
$\ket{\zeta} = \hat{S}(\zeta) \ket{0}$,
where the complex squeezing parameter is given by $\zeta = re^{2\mathrm{i}\theta}$, and the squeezing operator is given by $\hat{S}(\zeta) = \exp\left( \frac{1}{2} (\zeta^*)^2 \hat{b}^2 - \frac{1}{2} \zeta^2 (\hat{b}^\dagger)^2 \right)$, can be computed from the definition of the canonical positive-$\mathcal{P}$ distribution, as introduced in \cite{Olsen_2009}
\begin{equation}
P_\mathrm{sq, can}(\alpha, \tilde{\alpha}^*)
=\frac{e^{-\nu_x^2 /\left(e^{-r} \cosh r\right)}}{\sqrt{\pi e^{-r} \cosh r}} \frac{e^{-\nu_y^2 /\left(e^r \cosh r\right)}}{\sqrt{\pi e^r \cosh r}} \frac{e^{-|\delta|^2}}{\pi}.
\end{equation}
 This distribution can be sampled as follows:   $\alpha =e^{\mathrm{i} \theta} \nu+\delta, \quad
\tilde{\alpha} =e^{\mathrm{i} \theta} \nu-\delta$, where $ \delta=\frac{1}{\sqrt{2}}\left(q_1+\mathrm{i}q_2
\right),
 \nu=\nu_x+\mathrm{i}\nu_y=\sqrt{\frac{e^{-r} \cosh r}{2}} q_3
 +\mathrm{i} \sqrt{\frac{e^r \cosh r}{2}} q_4$, 
 and $q_{1}$, $q_{2}$, $q_{3}$, $q_{4}$
 are uncorrelated random Gaussian variables with zero mean and variance 1. In contrast to a coherent state, obtaining an accurate value of mean density or quadratures of a squeezed vacuum state requires averaging over stochastic samples.

The Schrödinger cat  state is defined as a quantum superposition of two coherent states with opposite phases
$ \ket{cat} = \mathcal{N}\left(\ket{\beta} + e^{\mathrm{i}\theta}\ket{-\beta} \right),
$ where the normalisation factor is given by $
\mathcal{N} = 1/{\sqrt{2\left(1+e^{-2|\beta|^2} \cos \theta\right)}}$.

For the Schrödinger cat state, the density matrix $ \hat{\rho}_{cat} $ is easily expressed in the coherent state basis with four elements proportional to $ \ket{\beta}\bra{\beta} $, $ \ket{\beta}\bra{-\beta} $, $ \ket{-\beta}\bra{\beta} $, and $ \ket{-\beta}\bra{-\beta} $. To determine the probability distribution value for $ \left(\alpha, \tilde{\alpha}^{*}\right) $, we compute two scalar products: $ \bra{\frac{1}{2}\left(\alpha + \tilde{\alpha}\right)}\ket{\beta}, \quad  \bra{\frac{1}{2}\left(\alpha + \tilde{\alpha}\right)}\ket{-\beta} $ and multiply them by the appropriate factors. For an arbitrary cat state, the canonical distribution can then be expressed as
\begin{align}
\begin{split}
P_{\mathrm{cat, can}}\left(\alpha, \tilde{\alpha}^*\right) = |\mathcal{N}|^{2}e^{-\frac{1}{2}\left(|\alpha|^{2} + |\tilde{\alpha}|^2\right)}e^{-|\beta|^{2}} 
\\
\times 
\left[ e^{\frac{\left(\alpha + \tilde{\alpha}\right)^{*}}{2}\beta} + e^{\mathrm{i}\theta}e^{-\frac{\left(\alpha + \tilde{\alpha}\right)^{*}}{2}\beta}\right]^{2}.
\label{p_can_cat}
\end{split}
\end{align}
We can use the above distribution to sample $\alpha$  and $\tilde{\alpha} ^*$, to construct the Schrödinger cat state.


It is important to note that, from a numerical perspective, each field is represented by complex variables. Therefore, the 
phase-space distribution under consideration is four-dimensional.
Sampling large cat states from this distribution may require $\sim10^5-10^6$ stochastic trajectories for accuracy.  The method results become exact in the limit $\mathcal{S}  \rightarrow \infty$. However, our results demonstrate that the proposed sampling method yields satisfactory accuracy for cat states that are feasible from an experimental perspective ($\langle \hat{n} \rangle \in(1,3)$). For a detailed numerical representation of many kinds of quantum states in the PPM, we refer to the works~\cite{Olsen_2009,Drummond20}. 

It is relevant to point out that the visualisation of quantum states via the positive $\mathcal{P}$ representation is non-trivial due to its four-dimensional structure, involving the real and imaginary parts of the stochastic variables $\alpha$ and $\tilde{\alpha}$. To improve interpretability, we present the $P$ distribution alongside the Wigner quasi-probability distribution, which offers a more standard and intuitive visualisation for single-mode quantum states. A comparison between the $\mathcal{W}(p,q)$ and $P(\alpha,\tilde{\alpha}^*)$ distributions is shown in Fig.~\ref{fig:fig8}.

\textit{II. Input Signal Injection}.  Following \cite{Ghosh_2019}, we assume that the input state is unidirectionally coupled to the system, which allows the use of the \textit{cascade formalism}, describing a one-way source-system interaction~\cite{Lopez_2016}. 
 In this approach, the input nodes are no longer integral to the reservoir. Therefore, equation~(\ref{Eq:1}) should be modified to introduce input source to reservoir coupling, with the source being as single-mode quantum state. Denoting input mode creation and annihilation operators as $\hat{s}$ and $\hat{s}^\dagger$, respectively, the master equation 
 takes the 
 form

\begin{align}
    \label{Eq:9}
\begin{split}
\dot{\hat{\rho}}\left(t\right)=-\mathrm{i}\left[{\hat{H}},\hat{\rho}(t)\right]+
\sum_{j=1}^N\frac{\gamma_j}{2}\mathcal{L}(\hat{b}_j) + \frac{\gamma_s\eta}{2}f(t)\mathcal{L}(\hat{s})\\ +\sum_{j}\sqrt{\gamma_{s}\gamma_{j}}\sqrt{f(t)}\text{W}_{j}^{\text{in}}([\hat{s}\hat{\rho},\hat{b}_j^\dagger]+[\hat{b}_j,\rho\hat{s}^\dagger]),
\end{split}
\end{align}
where 
$\gamma_s$ 
is the decay rate of the source mode. The function $ f(t)$ represents the temporal envelope of the input coupling, which is modelled as a Heaviside function. The vector $\text{W}_j^{\text{in}}$ denotes a set of random input weights, each sampled from a uniform distribution within the range $[0, 1]$. 
To guarantee that the photons from the source states that have entered the reservoir are being removed from the source, it is necessary that the parameter $\eta =  \sum_{j}\left(\text{W}_{j}^{\text{in}}\right)^{2}$. The presence of additional reservoir modes in the reservoir increases the effective dissipation rate of the source mode, as evidenced by the greater value of the parameter $\eta$.

Consequently, the system of Eqs. (\ref{eq:PP}) should be amended to incorporate the time dependence of the source cascade terms. Therefore, the stochastic equation of motion, in positive-$\mathcal{P}$ formalism and presence of cascade source-reservoir coupling, can be written as
\begin{align}
\begin{split}
\label{Eq:8}
&\frac{\partial \alpha_i}{\partial t} =\left. \frac{\partial \alpha_i}{\partial t}\right|_{R}- \sqrt{\gamma_{s}\gamma_{i}}\sqrt{f(t)}\text{W}_i^{\text{in}} s , \\
&\frac{\partial \tilde{\alpha}_i}{\partial t}=\left.\frac{\partial \tilde{\alpha}_i}{\partial t}\right|_R- \sqrt{\gamma_{s}\gamma_{i}}\sqrt{f(t)}\text{W}_i^{\text{in}} \tilde{s} , \\
&\frac{\partial s}{\partial t}=-f(t) \frac{ \gamma_s\eta}{2} s, \\
&\frac{\partial \tilde{s}}{\partial t}=-f(t) \frac{\gamma_s\eta}{2 } \tilde{s}.
\end{split}
\end{align}
In the above set of equations, functions $\frac{\partial \alpha_i}{\partial t}\vline_R$ and $\frac{\partial \tilde{\alpha}_i}{\partial t}\vline_R$ correspond to the right-hand sides (RHSs) of Eqs.~(\ref{eq:PP}). 
For detailed derivation of Eqs.~(\ref{Eq:8}), and details on numerical implementation, please refer to Appendix~\ref{sec:appB} and~\ref{appC}.

\textit{III. Output Interpretation}. The time-dependent average occupation of the reservoir nodes can be computed by stochastic averaging over realizations as
\begin{equation}
\langle \hat{n}_j\rangle(t)=\Re\langle(\alpha_{j}(t)\tilde{\alpha}_{j}^{*}(t))\rangle_S,
\end{equation}
and later integrated over a given time window to construct the $\mathbf{n}$ vector (as detailed in Sec.~\ref{sec:modelA}) used to map quantum information to classical observables for performing the given quantum machine learning task. 
\section{Results}
 To validate the performance of a large-scale bosonic QRNN using the positive-$\mathcal{P}$ method, we examine the QSC and QFP tasks.
 
\begin{figure}
\includegraphics[width=0.76\linewidth]{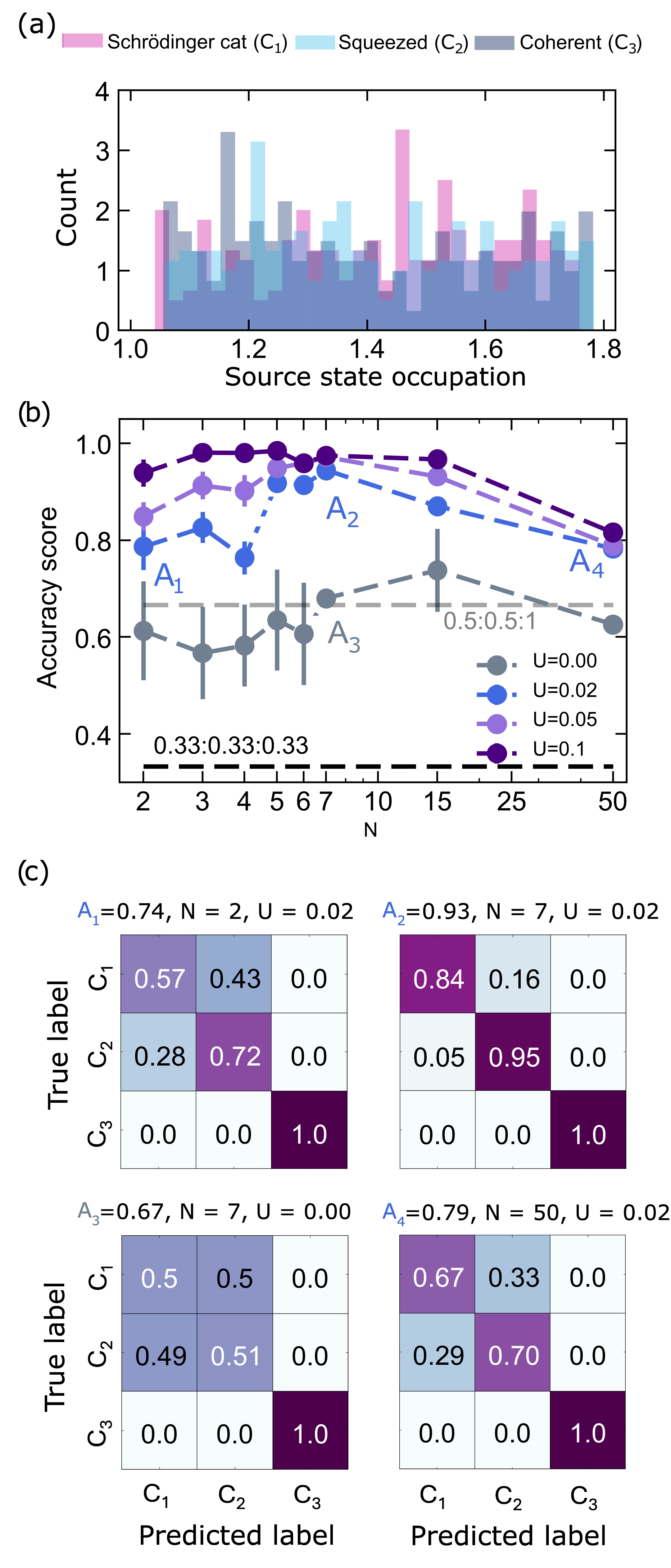}
\caption{ \textbf{Quantum state recognition with quantum reservoirs.} (a) Distribution of average occupation $\left\langle \hat{n}\right\rangle$ of source quantum states used during the training phase, illustrating the sampling over the quantum state parameter space. (b) Average (over three runs) test accuracy as a function of reservoir size for four different nonlinearity values, depicted on the plot. Reservoirs examined in this study were composed of $N \in \{2,3,4,5,6,7,15,50\}$ nodes.
Dashed lines indicate baseline accuracies of classification for random and partially random guesses. The black line corresponds to uniform guessing among three classes with ratio  $(0.33:0.33:0.33)$, and the grey line denotes uniform guessing between classes with ratio $(0.5:0.5:1)$, meaning the third class is predicted correctly. 
Error bars represent the standard deviation across the three independent training runs. (c) Confusion matrices for representative quantum reservoirs, each corresponding to the test accuracy $A_i$, 
highlighted in the panel above. The matrices illustrate classification performance during inference. The labels $C_1$, $C_2$, and $C_3$ refer to Schrödinger cat states, squeezed states, and coherent states, respectively.}
    \label{fig:fig3x}
\end{figure}

\subsection{Quantum state classification}
\label{sec:class}
This task focuses on identifying the quantum state optically injected into the reservoir, assuming that the input quantum states belong to one of three predefined categories: Schrödinger-cat, squeezed, and coherent states, as illustrated in Fig.~\ref{fig:fig1}b.

To perform the quantum state classification task, we construct training $\mathbf{D}_{\text{train}} = \{(\mathbf{x}_i, \mathbf{y}_i)\}_{i=1}^{N{\text{train}}}$ and testing datasets $\mathbf{D}_{\text{test}} = \{(\mathbf{x}_i, \mathbf{y}_i)\}_{i=1}^{N{\text{test}}}$, containing $N_{\text{train}}$ and $N_{\text{test}}$ samples, respectively. Each feature vector $\mathbf{x}_i$ encodes the stochastic configuration used during sampling of a specific input state, while each label $\mathbf{y}_i$ is represented using one-hot encoding, where each class is denoted by a binary vector, as shown in Fig.~\ref{fig:fig1}a.

To ensure that the prepared quantum states effectively validate the reservoir's capabilities in the classification task, we generate quantum states with average occupation uniformly distributed over a specified range with high interclass overlap (see Fig.~\ref{fig:fig3x}a) 
This approach mitigates the risk that the reservoir learns to detect only the average occupation rather than the distinctive quantum correlations encoded in the states' density matrices. The state parameters (as defined in \ref{QRCstates}) are drawn from uniform distributions over the following ranges. For Schrödinger-cat states, where $\beta = |\beta|e^{\mathrm{i}\varphi}$, $|\beta| \in (1.12, 1.38)$ and $\varphi \in \left(0, \frac{\pi}{2}\right)$, for squeezed states, $r \in (0.9, 1.1)$ and $\theta \in \left(0, \frac{\pi}{2}\right)$, and for coherent states, where $\beta = |\beta|e^{\mathrm{i}\varphi}$, $|\beta| \in (1.03, 1.34)$ and $\varphi \in \left(0, \frac{\pi}{2}\right)$.

Using the PPM to model large-scale quantum dynamics, we examine the dependence of classification accuracy on reservoir size for four values of Kerr nonlinearity (identical for each node), $U/\gamma = {0.0, 0.02, 0.05, 0.1}$, where $\forall_{i} \gamma_{i} = \gamma_{s} = 1 = \gamma$, and $\Delta_{i}$ is uniformly distributed from  $\Delta_{j}\in(0,0.1\gamma)$. The values of the hopping amplitudes $J_{ij}$ were chosen from a uniform distribution in the range $J_{ij}\in(-1,1)$, normalised by the spectral radius (largest modulus of eigenvalues) ~\cite{Ghosh_2019}. As shown in Fig.~\ref{fig:fig3x}b, for $U = 0.02$, the accuracy (validated on the testing dataset) increases with the number of lattice sites, peaking at $N = 7$, with optimal values of $N \in \{5, 6, 7\}$.
Beyond this range, the performance decreases. 
As $N$ increases,  photons are distributed over more modes, lowering the photon count per mode and thus reducing the effective nonlinearity. For large $N$, the reservoir dynamics approaches linear behaviour, diminishing the system's ability to capture quantum correlations essential for accurate classification. This decline becomes noticeable at $N = 15$ and is pronounced at $N = 50$. We can also see that raising $U$ brings back good performance  e.g.for $N=15$. Therefore, this degradation is attributed to the reduction in \textit{effective nonlinearity}, resulting from the Kerr nonlinearity and the average number of photons per reservoir node.

To further analyse the results of quantum states classification, we compute confusion matrices (based on the testing dataset) for four reservoir configurations with selected values of nonlinearity and reservoir size, as shown in Fig.~\ref{fig:fig3x}c.
Considering the properties of coherent states (denoted as $C_3$), we observe that, even in the absence of nonlinear interactions, the reservoir is capable of distinguishing coherent states from Schrödinger-cat ($C_1$) and squeezed states ($C_2$). This indicates that the correlations of reservoir average density dynamics are sufficient for identifying coherent states, without requiring nonlinear effects. A reference line for which only coherent states are correctly classified, corresponding to the diagonal $(0.5, 0.5, 1)$, is indicated in panel (b) by the grey dashed line. The corresponding total prediction score (defined as the fraction of states that have been correctly classified) 
is $A = 0.66$, significantly above the random baseline of $A = 0.33$ (black dashed line).

As our simulations confirmed, increasing the Kerr nonlinearity improves classification accuracy and broadens the range of reservoir sizes over which high performance is maintained. Reservoirs with higher effective nonlinearity retain classification accuracy even at larger sizes, whereas those with weaker nonlinearity exhibit accuracy degradation for smaller $N$. Similar trends are observed in the quantum feature prediction task, discussed in the following subsection.

\subsection{Quantum state feature prediction}
\label{sec:QSFP}

This task focuses on reconstructing the complex squeezing parameter of a squeezed vacuum state by predicting real and imaginary components of the parameter. As in the QSC task, we investigate how the reservoir size and Kerr nonlinearity affect the reservoir's properties for quantum feature prediction. To quantify prediction precision, the mean-squared error (MSE), defined as
\begin{equation}
\text{MSE} = \frac{1}{n}\sum_{i=1}^n\left| \zeta_{i} - \bar{\zeta}_i \right|^{2},
\label{mse}
\end{equation}
was used, where $n$ is the number of states in the validation dataset (either testing or training), and $\zeta_i$ and $\bar{\zeta}_i$ are the predicted and target parameters $\zeta_i = r_ie^{\mathrm{i}2\theta_i}$, respectively. The MSE metric quantifies the average squared distance in the complex plane between predicted and target parameters.

\begin{figure}[t!]
\includegraphics[width=0.95\linewidth]{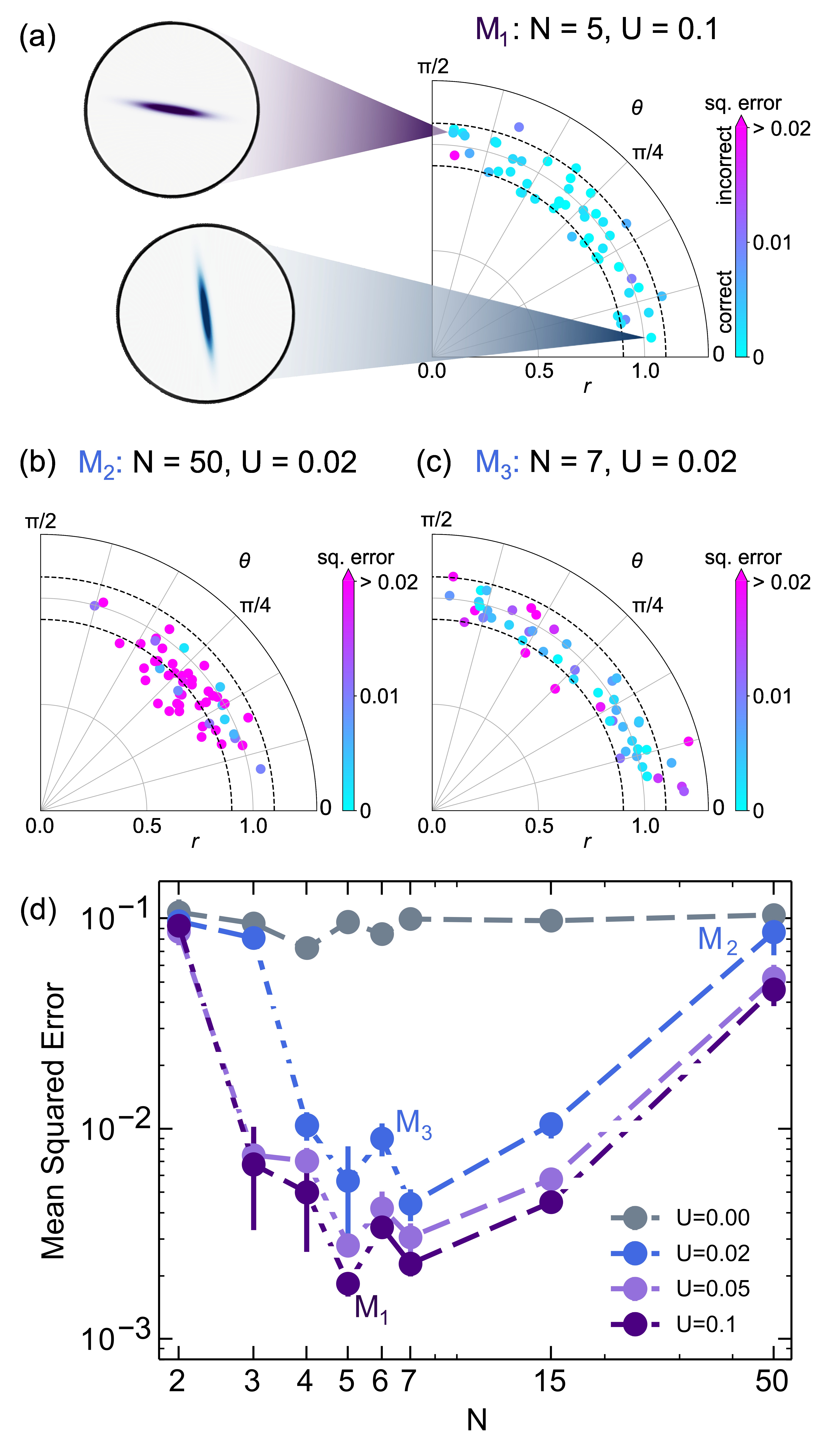}
\caption{\textbf{Prediction of squeezed vacuum state parameter.} 
(a)--(c) Polar plots illustrating the precision of squeezing parameter prediction using quantum reservoirs of increasing size: 
(a) 5-mode reservoir with Kerr nonlinearity $U = 0.1$, 
(b) 50-mode reservoir with $U = 0.02$, and
(c) 7-mode reservoir with $U = 0.02$. 
Each point represents a predicted squeezing parameter.  Subsets in panel (a) illustrate Wigner functions of two squeezed states with different squeezing parameters. 
(d) Average, mean-squared error of the predicted squeezing parameter $\zeta$ (real and imaginary components) as a function of reservoir size, evaluated for different Kerr nonlinearities. Error bars correspond to the standard deviation computed from five distinct training runs. 
}
    \label{fig:fig3}
\end{figure}

Although the training process yields predictions for the real and imaginary components of $\zeta_i$, it is more informative to visualise the results in polar coordinates, where the radial and angular components correspond to the squeezing amplitude $r_i$ and phase $\theta_i$, respectively. Figures~\ref{fig:fig3}a--c show the results for various Kerr nonlinearities and reservoir sizes, with each point representing a predicted squeezing parameter. To enhance interpretability of the results, we plot each of the predicted points with a colour scale resembling the error of prediction measured as the squared distance on the complex between the predicted and target point $\left| \zeta_{i} - \bar{\zeta}_i \right|^{2}$. The distribution of squeezed state parameters used during training is the same as in Section~\ref{sec:class}. 

Figure~\ref{fig:fig3}d illustrates the MSE of the predicted squeezing parameter as a function of reservoir size for various Kerr nonlinearities. For all reservoir sizes, the error decreases with increasing nonlinearity, consistent with the expected behaviour of reservoir networks. Prediction accuracy improves with increasing reservoir size initially, but this trend reverses for large reservoirs (e.g., $N = 15$ and $N = 50$). As observed in the classification task, this performance decline stems from the reduction of \textit{effective nonlinearity}. Specifically, reservoir performance decreases as the average photon number is distributed over more reservoir modes. In such cases, the reservoir fails to learn the proper distribution of squeezing parameters, producing random predictions biased toward the average of the training samples. This effect is illustrated in Fig.~\ref{fig:fig3}b 
as a cloud of magenta 
points centered around $\pi/4$, for reservoir size and nonlinearity set to $N=50$ and $U=0.02$.

\section{Discussion}
The reservoir computing approach enhances scalability and significantly reduces the computational complexity of training quantum neural networks, eliminating the need for advanced optimisation methods such as backpropagation~\cite{Pan_2023}. However, we now can conclude by demonstration 
that achieving high reservoir performance requires balance between nonlinearity and reservoir size.
Our results also  demonstrate that the positive-$\mathcal{P}$ representation is suitable for simulating large-scale quantum reservoirs in neuromorphic quantum computing. To objectively assess its broader applicability, we now discuss the known limitations of this approach, particularly regarding simulation stability and the initialisation of input quantum states in phase space.

\subsection{Stability of the method and range of applicability}
The positive-$\mathcal{P}$ method can become unstable in closed systems due to the effect of \textit{noise amplification}, present during numerical integration of stochastic differential equations \cite{Gilchrist97,Deuar_2002,Deuar_2006}. However, in open quantum systems, sufficiently large dissipation can stabilise the stochastic trajectories~\cite{Deuar_2021}. The stability regime in such systems is primarily governed by the ratio of nonlinear interaction strength to dissipation. A detailed numerical analysis of stability—based on the fraction of convergent trajectories during numerical simulations of a bosonic reservoir as a function of Kerr-like nonlinearity $U$, pump amplitude $F$, and dissipation rate $\gamma$-is provided in Appendix~\ref{app: stab}. Under typical operational parameters, the method’s intrinsic stability accurately describes the optical quantum neural network.

\subsection{Initialisation of input quantum states}
As was detailed in the previous section (see Sec. \ref{QRCstates}), the positive-$\mathcal{P}$ method requires a well-defined phase-space distribution for initialising quantum states. The process of initialisation is straightforward for Gaussian states, such as coherent and thermal states, but becomes challenging for non-Gaussian or entangled states, where the corresponding positive-$\mathcal{P}$ distribution may lack a tractable analytical form or be challenging to sample numerically. Consequently, input initialisation may limit the class of quantum states that can be reliably injected into the reservoir. Nonetheless, as shown in Sec.~\ref{QRCstates}, the method remains capable of simulating a wide range of practically relevant quantum states~\cite{Olsen_2009}.

\begin{figure}[t!]
\includegraphics[width=1\linewidth]{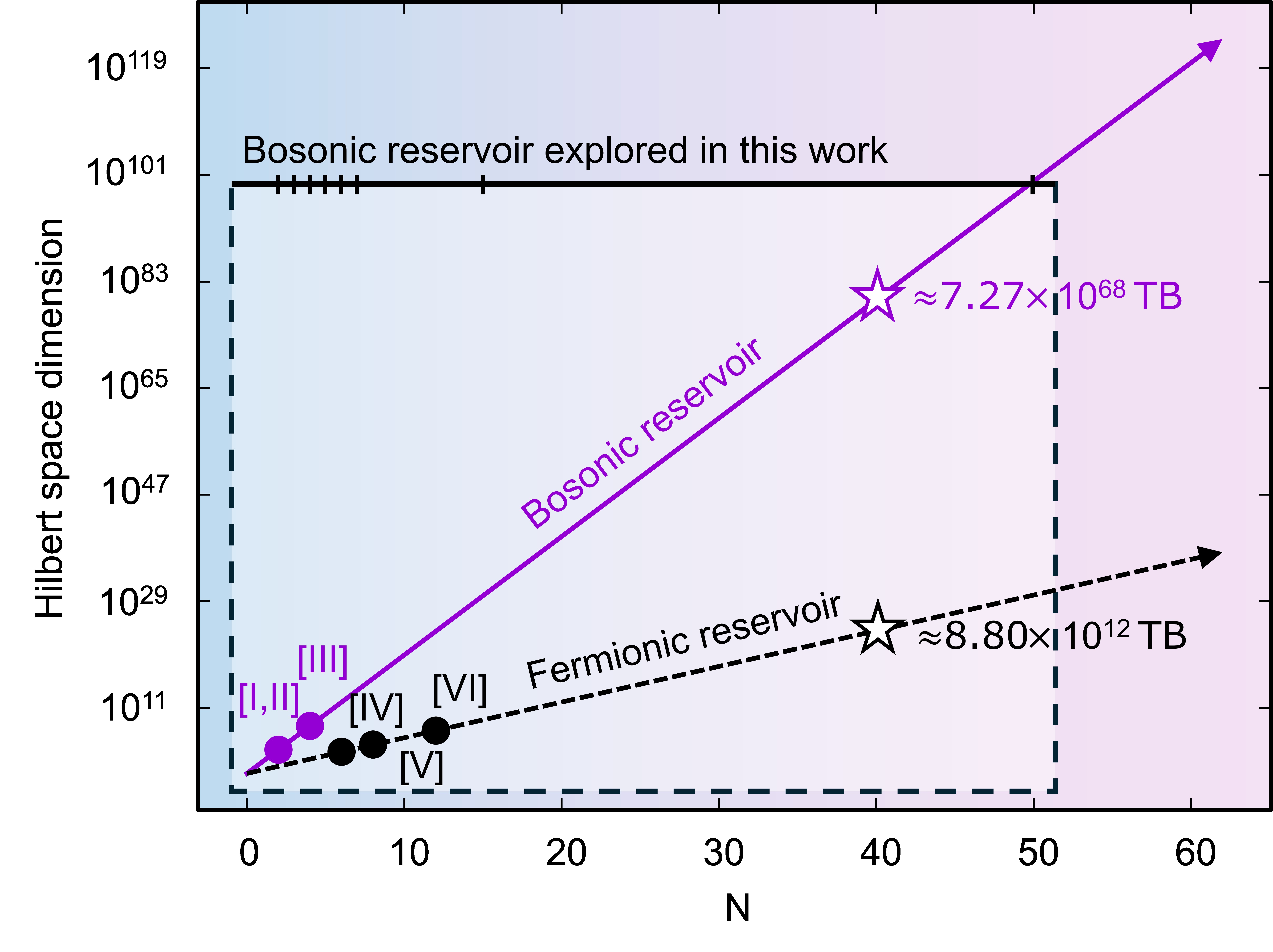}
    \caption{\textbf{Complexity of QRC with fermionic and bosonic reservoirs.}
This figure illustrates the computational complexity of QRC, showing how the Hilbert space dimension of the reservoir scales with the number of nodes for fermionic (black dashed line) and bosonic (violet solid line) reservoirs. In the bosonic case, the local occupation number is truncated to a maximum of 10 photons per site ($d = 10$). Violet and black dots indicate quantum neuromorphic systems reported in the literature, for bosonic and fermionic systems, respectively. Labels 
[I]-[VI] indicate the works 
~\cite{Ghosh_2019},~\cite{Türeci_2021},~\cite{Xu_2023},~\cite{Ghosh_2019},~\cite{Ghosh_2021_rec},~\cite{Suzuki_2022}. White stars highlight configurations where the density matrix size-assuming 64-bit floating-point precision (float 64)-reaches  $\sim 10^{68}$ TB and $\sim 10^{12}$ TB, respectively. The grey shaded region denotes the range of reservoir sizes explored in this study.}
    \label{fig:fig4}
\end{figure}

\subsection{Comparison with other approaches}
Alternative methods offer complementary advantages but also notable limitations. The bosonic truncated Wigner approximation (TWA)~\cite{Steel_1998,Sinatra_2002} achieves good 
numerical stability in dissipative systems \cite{Deuar_2021,Orso25} but does this by neglecting higher-order quantum terms,  
and therefore fails to capture full quantum correlations outside of the semi-classical regime attained with high enough occupations. Matrix product state (MPS) and tensor network techniques efficiently capture entanglement in low-dimensional systems but usually scale poorly in higher spatial dimensions or mode occupations; moreover they do not give access to single realizations. Other time-dependent variational approaches can sometimes bring insight, especially when based upon quantum trajectories \cite{Verstraelen2020, Daley04032014}, but lack the full precision and ability to track arbitrary correlations of the PPM. Quantum Monte Carlo methods (QMCM)~\cite{Knight_1998} are effective for equilibrium and imaginary-time simulations but suffer from severe sign problems during real-time evolution~\cite{Meng_2024}. In contrast, the positive-$\mathcal{P}$ method, when outside the noise amplification instability regime, offers a key advantage. Compared to TWA, MPS, and QMCM, the  positive-$\mathcal{P}$ method provides asymptotically exact simulations of quantum dynamics in large, multimode, interacting bosonic systems without truncating the Hilbert space or imposing spatial dimensional constraints. This makes it particularly suitable for modelling high-dimensional quantum fields and applications such as quantum optical neural networks, where scalability and fidelity are paramount.

Figure~\ref{fig:fig4} illustrates the computational complexity of quantum reservoir computing using bosonic and fermionic reservoirs. The complexity of simulation is reflected by the size of the density matrix, which scales with the dimension of the computational Hilbert space as $\sim d^{2N}$, where $d$ is the cut-off of Hilbert space dimension and $N$ is the number of reservoir modes. This analysis considers only the reservoir density matrix and excludes contributions from the input state space, which would further increase complexity, as discussed in Eq.~(\ref{eq:1}). Due to more favourable Hilbert space scaling, most bosonic QRC studies have focused on small systems with 2–5 bosonic modes, while fermionic reservoirs have been explored in configurations involving up to 10 modes. The violet and black dashed lines show the Hilbert space scaling for fermionic and bosonic systems. The grey region highlights the parameter range examined in this work. Notably, the use of the positive-$\mathcal{P}$ method enables the simulation of a reservoir whose density matrix, under standard methods (e.g., master equation approaches) with a truncation at $d=10$, would require over $ 10^{37}$ TB of memory. This illustrates the impracticality of conventional density-matrix-based simulations for such large-scale quantum systems.

\section{Conclusions} 
In conclusion, this work presents a scalable framework for simulating quantum bosonic neural networks using the \textit{phase-space} PPM. Our study provides the first comprehensive theoretical validation of intermediate-scale bosonic reservoirs with Kerr-like nonlinearity, particularly for quantum state classification and feature prediction. Interestingly, we observe a counter-intuitive phenomenon: unlike in typical classical systems, increasing the reservoir size beyond a certain threshold decreases neural network performance. We show that reservoirs with five to seven nodes achieve optimal performance for the analysed quantum machine learning tasks. This phenomenon arises from the distribution of information from the input quantum state across all reservoir nodes, which effectively reduces the system's nonlinearity in larger reservoirs. Our findings provide a theoretical foundation for designing quantum optical neuromorphic devices, with significant implications for practical applications in quantum computing and quantum machine learning.

\section{Acknowledgments}

A. O. acknowledges the National Science Center, Poland, project No. 2024/52/C/ST3/00324.
M. M. acknowledge support from National Science Center, Poland (PL), Grant No.~2021/43/B/ST3/00752.
This work was financed by the European Union EIC Pathfinder Challenges project ``Quantum Optical Networks based on Exciton-polaritons'' (Q-ONE, Id: 101115575). We are thankful to all partners of the Q-One project for the supportive discussion, in particular to Daniele Sanvitto, Vincenzo Ardizzone, Alberto Bramati, Mathias Van Regemortel, and 
Thomas Van Vaerenbergh.
\section{Data Availability}
The data that support the findings of this article are openly available \cite{Swierczewski2025}.
\section{APPENDIX}
\appendix{}
\section{Learning process}
\label{sec:appA}
\begin{figure}[h!]
    \centering
    \includegraphics[width=\linewidth]{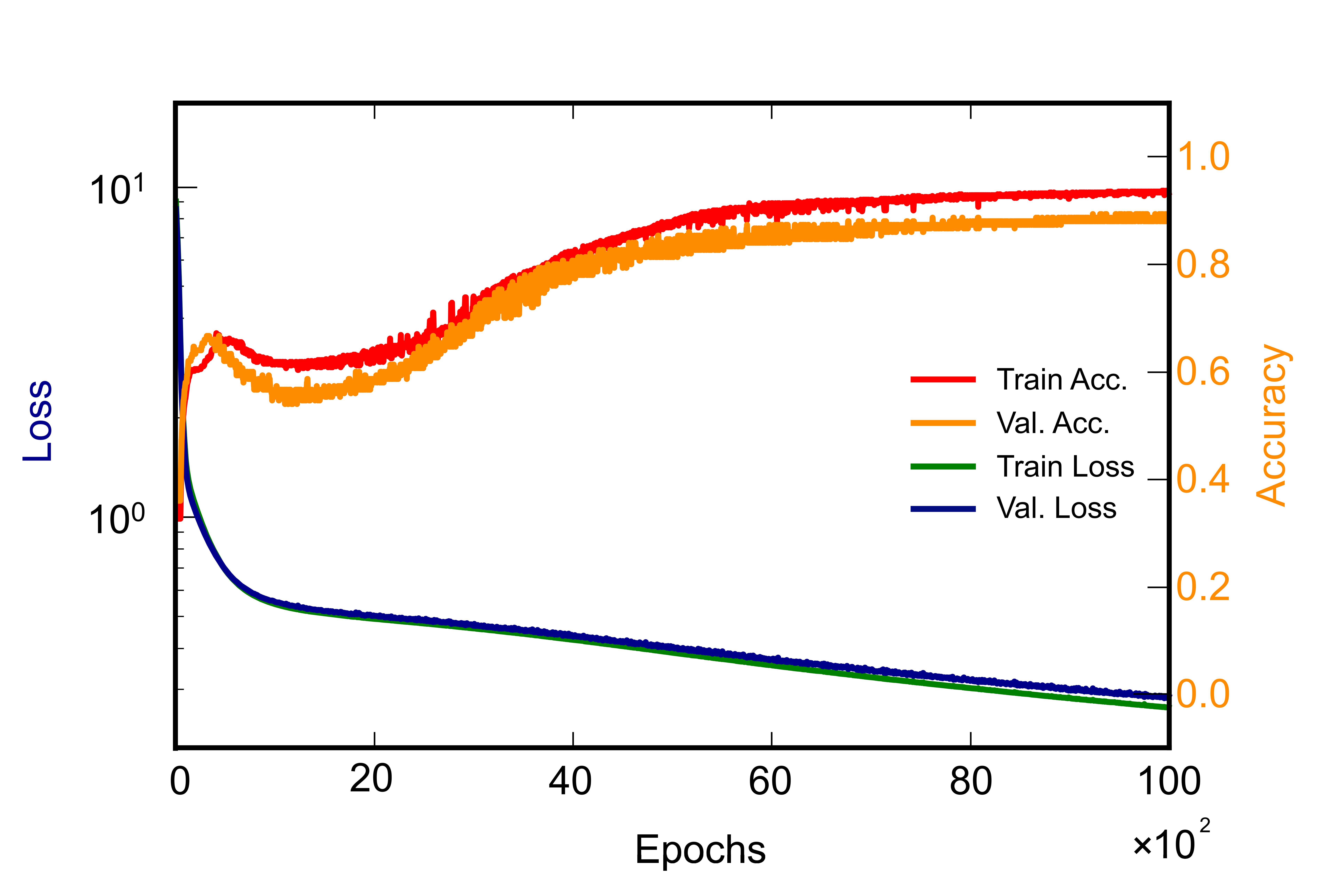}
    \caption{Training process for a reservoir of size N = 7 and nonlinearity U = 0.02. Loss function (categorical cross-entropy) has been plotted as a function of epochs for the training (green) and testing (blue) data. A second metric, the categorical accuracy, has been plotted for the training (red) and validation (orange) sets. As the network is trained, along with the decrease of loss function, an increase of accuracy is observed until saturation around 60-80 epochs.}
    \label{fig:training}
\end{figure}

In the classification task, we employed the softmax function to compute the probability distribution over all considered classes ($j = 1, 2, \ldots, K=3$). The softmax function is defined as follows
\begin{equation}
P(y = j | \mathbf{n}) = \frac{\exp(\mathbf{W}_j \mathbf{n} + b_j)}{\sum_{k=1}^K \exp(\mathbf{W}_k \mathbf{n} + b_k)},    
\end{equation}
where $\mathbf{W}_j$ is the $j$-th row of the weight matrix $\mathbf{W}$, and  
$b_j$ is the $j$-th bias term. The matrix $\textbf{W}$ is calculated by minimising the loss function using optimisation techniques such as gradient descent, which was used here. In the multi-class classification task ($K>2$), choosing the categorical cross-entropy as the loss function, given by the following equation, is the most common choice \cite{dawid_2023} and was also employed in our work for quantum state classification
\begin{equation}
L = -\frac{1}{N} \sum_{i=1}^N \sum_{j=1}^K y_{i,j} \log P(y = j \mid \mathbf{n}_i).
\end{equation}
 The training was performed using TensorFlow, using the \textit{Adam} optimiser, a stochastic gradient descent model. From every class of quantum states, 200 states have been chosen for the training dataset. The validation set has been composed of 50 states per class. The training has been performed over $10\cdot 10^3$ epochs with a learning rate of $dl =5\cdot10^{-4}$ with one batch containing all 600 states. This choice of parameters ensured convergence, and an example of the training process for the classification task has been shown in Figure~(\ref{fig:training}). 

For the feature prediction task, where the squeezing parameter was being predicted, the MSE has been used as the loss function, a standard choice in regression tasks. The formula for the MSE is given by equation (\ref{mse}) in the main text. For this task, for each reservoir size and nonlinearity, 200 states have been used for the training process with a batch of 150 states, and 50 states were used for validation. The training has been performed over $25\cdot 10^3$ epochs with a learning rate $dl =5\cdot10^{-4}$. Once again, the \textit{Adam} optimiser has been employed for the training.

\section{Numerical implementation}
\label{sec:appB}
For the numerical treatment of the  positive-$\mathcal{P}$ equations governing the reservoir dynamics (see Eq. (\ref{Eq:8})), we have chosen the semi-implicit midpoint algorithm \cite{DRUMMOND1983211}. The algorithm's applicability for solving stochastic differential equations in the positive-$\mathcal{P}$ framework has been studied in \cite{Deuar2021multitime}, where a detailed description of the method and its applicability has been provided. Here, we provide the algorithm used to numerically solve Eq.~(\ref{Eq:8}).

The positive-$\mathcal{P}$ equations are a set of Ito stochastic equations characterised by the deterministic \textit{drift} terms $A_j(\vec{v}, t)$ and the stochastic \textit{noise} terms $X_j(\vec{v}, t)$. The general form of such equations is\\
\begin{equation}
    \frac{d v_j}{d t}=\mathcal{D}_j(\vec{v}, t)=A_j(\vec{v}, t)+X_j(\vec{v}, t),
\end{equation}
where  $\vec{v}$ is a vector of all the sampled fields (reservoir and source). First, we split the evolution into terms which will be treated exponentially ($\mathcal{D}^E$) in the algorithm and the rest of the terms ($\mathcal{D}^{R}$). This separates the evolution equations into
\begin{equation}
    \frac{d v_j}{d \tau}=\mathcal{D}_j^E v_j+\mathcal{D}_j^R.
\end{equation}
This separation for equations (\ref{Eq:8}) has been done in the following form, where the evolution equations for the reservoir modes consist of both the exponential and rest terms, however, the source mode evolution is treated fully exponentially
\begin{align}
    &\mathcal{D}_{\alpha_{j}}^{E} = \mathrm{i}\Delta_j - \mathrm{i}U \alpha_j\tilde{\alpha}_j^* - \frac{\gamma_j}{2} 
     + \sqrt{-\mathrm{i}U}\xi_j(t) + \mathrm{i}\frac{U}{2}, \label{exp1}\\
    &\mathcal{D}_{\alpha_{j}}^{R} = - \mathrm{i}F_j  + \mathrm{i}\sum_k J_{kj}\alpha_k - \sqrt{\gamma_{s}\gamma_{j}}\sqrt{f(t)}\,\text{W}_j^{\text{in}} s,\\
    &\mathcal{D}_{\tilde{\alpha}_{j}}^{E} = \mathrm{i}\Delta_j - \mathrm{i}U \alpha_j\tilde{\alpha}_j^* - \frac{\gamma_j}{2}
     + \sqrt{-\mathrm{i}U}\tilde{\xi}_j(t) + \mathrm{i}\frac{U}{2}, \label{exp2} \\
    &\mathcal{D}_{\tilde{\alpha}_{j}}^{R} = - \mathrm{i}F_j  + \mathrm{i}\sum_k J_{kj}\tilde{\alpha}_k - \sqrt{\gamma_{s}\gamma_{j}}\sqrt{f(t)}\,\text{W}_j^{\text{in}} \tilde{s},\\
    &\mathcal{D}_{s}^{E}  =-f(t) \frac{\eta\gamma_s}{2 },\\
    &\mathcal{D}_{\tilde{s}}^{E}  =-f(t) \frac{\eta\gamma_s}{2 }.
\end{align}
The final terms of equations (\ref{exp1}) and (\ref{exp2})  
autocorrelation corrections \cite{Deuar2021multitime} tailored to the time-stepping algorithm (sometimes dubbed ``Stratonovich corrections'', but their presence is not limited to Ito/Stratonovich calculus conversions). The semi-implicit midpoint algorithm used to integrate the above evolution equations is given by the following form
$$
\begin{aligned}
& v_j^{\text {step }}\left[v_j^{(0)}, \mathcal{D}\left(\vec{v}^{(0)}, t\right), \tau\right]= \\
& \quad v_j^{(0)} e^{\tau \mathcal{D}_j^E\left(\vec{v}^{(0)}, t\right)}+\left(e^{\tau \mathcal{D}_j^E\left(\vec{v}^{(0)}, t\right)}-1\right) \frac{\mathcal{D}_j^R\left(\vec{v}^{(0)}, t\right)}{\mathcal{D}_j^E\left(\vec{v}^{(0)}, t\right)}. \\
\end{aligned}
$$
The parameter $\tau$ in the above equations is the simulation time-step, and for the dynamics computed in this work was equal to $\tau = 0.05$, which has proven through trial and error to be sufficiently small to ensure stability of the method. The maximum simulation time was T = 25, which corresponded to all excitation of the reservoir modes by source modes dying out, leading back to a stationary state. 

\section{Derivation of cascade terms using positive-$\mathcal{P}$ representation} 
\label{appC}
To derive the equations of motion using the positive-$\mathcal{P}$ representation for a bosonic lattice (\textit{reservoir}) composed of Kerr oscillators, which is coupled via cascade coupling to an external quantum source as described by Eq. (\ref{Eq:9}), we extend the theoretical framework developed in \cite{Deuar_2021}. To guide the derivation, we first recall the fundamental properties of the coherent state kernel, writing directly the action of the $\hat{\Lambda}_i$ operator on the creation and annihilation operators, as detailed below

\begin{align}
& \hat{b}_i \hat{\Lambda}_i=\alpha_i \hat{\Lambda}_i, \nonumber \\
& \hat{b}_i^{\dagger} \hat{\Lambda}_i=\left(\tilde{\alpha}_i^*+\frac{\partial}{\partial \alpha_i}\right) \hat{\Lambda}_i, \nonumber \\
& \hat{\Lambda}_i \hat{b}_i=\left(\alpha_i+\frac{\partial}{\partial \tilde{\alpha}_i^*}\right) \hat{\Lambda}_i, \nonumber \\
& \hat{\Lambda}_i \hat{b}_i^{\dagger}=\tilde{\alpha}_i^* \hat{\Lambda}_i. \nonumber
\end{align}

To derive the stochastic equations one proceeds as follows: First, we calculate the terms arising from the substitution of the density matrix in $P$ representation, described by Eq.~(\ref{eq:dminP}), into the Lindblad master equation~(\ref{Eq:9}). This procedure leads to an integral equation 
\begin{align}
& \int d^{4 M} \vec{v} \widehat{\Lambda} \frac{\partial P}{\partial t} \nonumber \\
& \quad=\int d^{4 M} \vec{v} P\left\{\sum_v A_v(\vec{v}) \frac{\partial}{\partial v}+\sum_{v v^{\prime}} \frac{D_{v v^{\prime}}(\vec{v})}{2} \frac{\partial^2}{\partial v \partial v^{\prime}}\right\} \widehat{\Lambda}, \label{eq:integral1}
\end{align}
where $\vec{v}$ is a vector of all the sampled fields (reservoir and source) and $v$, $v'$ are components of this vector. After integrating by parts, we obtain an integral containing derivatives of $P$ instead of $\hat{\Lambda}$ 
\begin{align} & \int d^{4 M} \vec{v} \widehat{\Lambda} \frac{\partial P}{\partial t} \nonumber \\ & \quad=\int d^{4 M} \vec{v} \widehat{\Lambda}\left\{-\frac{\partial}{\partial v} \sum_v A_v(\vec{v})+\sum_{v v^{\prime}} \frac{\partial^2}{\partial v^{\prime} \partial v} \frac{D_{v v^{\prime}}(\vec{v})}{2}\right\} P. \label{eq:integral2}
\end{align}
The equation under the integrals is a Fokker-Planck equation corresponding to the initial master equation.
To ensure readability, we divided the right-hand side (RHS) of equation~(\ref{Eq:9}) into seven parts ($ \text{RHS}=\sum_{i=1}^{7}{\mathcal{T}_i}$). Each of these parts 
corresponds to a term, denoted as $\mathcal{P}_i$, on the right-hand side of equation~(\ref{eq:integral1}). 

First, we analyse terms proportional to the \textit{detuning} $\Delta_i$, coming from the coherent evolution of the Hamiltonian Eq.~(\ref{Eq:2}). Taking into account that  $\left[{\hat{H}},\hat{\rho}\right]={\hat{H}}\hat{\rho}-\hat{\rho}{\hat{H}}$
\begin{align}
 &{\mathcal{T}_{1}}= -\mathrm{i} \sum_i-\Delta_i\left[\hat{b}_i^{\dagger} \hat{b}_i \rho-\rho \hat{b}_i^{\dagger} \hat{b}_i\right],
\end{align}
which, after a straightforward substitution using the properties of the coherent state kernel, can be mapped into the following equation in phase-space representation  
\begin{align}
\mathcal{P}_1=  \int d^{4 M} \vec{\alpha } P \left[-\mathrm{i} \sum_i-\Delta_i\left(\alpha_i \frac{\partial}{\partial \alpha_i}-\tilde{\alpha}_i^* \frac{\partial}{\partial \tilde{\alpha}_i^*}\right)\right] \hat{\Lambda}. \nonumber
\end{align} Secondly, we consider a term $\mathcal{T}_2$ that corresponds to the Kerr nonlinearity of reservoir sites $U$ 
\begin{align}
&{\mathcal{T}_{2}}= -\mathrm{i} \sum_i \frac{U}{2}\left(\hat{b}_i^{\dagger} \hat{b}_i^{\dagger} \hat{b}_i \hat{b}_i \rho-\rho \hat{b}_i^{\dagger} \hat{b}_i^{\dagger} \hat{b}_i \hat{b}_i\right). \end{align}
Compared to $\mathcal{T}_1$ above, the equation contains high-order terms. Such terms induce mutual interactions between the $\alpha$ and $\tilde{\alpha}$. Therefore the $\mathcal{P}_2$ takes following form
\begin{align}
\nonumber 
\mathcal{P}_2=\int d^{4 M} \vec{\alpha } P\left\{-\mathrm{i} \sum_i \frac{U}{2}\hat{A}(\alpha,\tilde{\alpha}^*)\right\} \hat{\Lambda},\nonumber
\end{align}
where $\hat{A}$ is given by:
$
\hat{A}(\alpha,\tilde{\alpha}^*)=2 \alpha_i^2 \tilde{\alpha}_i^* \frac{\partial}{\partial \alpha_i}-2 \tilde{\alpha_i}^{*2} \alpha_i \frac{\partial}{\partial \tilde{\alpha_i}^*}+\alpha_i^2 \frac{\partial^2}{\partial \alpha_i^2}-\tilde{\alpha}_i^{*2} \frac{\partial^2}{\partial \tilde{\alpha}_i^{*2}}.
$ The third term describes the gain in the studied system due to pumping via an external coherent laser source $F_i$. Taking into account that, in general, the pumping can be completed, we consider the following equation  
\begin{align}
&{\mathcal{T}_{3}}=-\mathrm{i} \sum_i {\color{black}{\bigg[}}F_i\left(\hat{b}_i^{\dagger} \rho-\rho \hat{b}_i^{\dagger}\right)+F_i^*\left(\hat{b}_i \rho-\rho \hat{b}_i\right){\color{black}{\bigg].}}
\end{align}
This equation can be immediately mapped to the corresponding representation, taking the form 
\begin{align}
\nonumber
\mathcal{P}_{3}=\int d^{4 M} \vec{\alpha } P \left(-\mathrm{i} \sum_i F_i \frac{\partial}{\partial \alpha_i}-F_i^* \frac{\partial}{\partial \tilde{\alpha_{i}}^{*}} \right)\hat{\Lambda}.
\end{align}
The fourth term describes the coupling between the reservoir nodes, where the coupling strength is determined by the coupling matrix element $J_{ij}$, following the equation
\begin{align}
\mathcal{T}_{4} =   \mathrm{i} \sum_{\left\langle i j\right\rangle} \left[ J_{i j}\left(\hat{b}_j^{\dagger} \hat{b}_i \rho-\rho \hat{b}_{j}^{\dagger} \hat{b}_i\right)+J_{i j}^*\left(\hat{b}_i^{\dagger} \hat{b}_j \rho-\rho \hat{b}_{i}^{\dagger} \hat{b}_j\right)\right], \nonumber
\end{align}
where $\left\langle i j\right\rangle$ denotes the connections between modes $i$ and $j$.
This equation can be mapped to the equation $\mathcal{P}_{4}$, which couples the fields $\alpha_{i}$, $\alpha_{j}$ corresponding to different reservoir modes
\begin{align}
\mathcal{P}_{4} = \int d^{4 M} \vec{\alpha } P\left[ \mathrm{i} \sum_{\left\langle i j\right\rangle} J_{i j} \hat{C}_{ij}+J_{i j}^* \hat{C}_{ji}\right]\hat{\Lambda}, \nonumber
\end{align}
where $\hat{C}_{ij}$ = $\left(\alpha_i \frac{\partial}{\partial \alpha_j}-\tilde{\alpha}_j^* \frac{\partial}{\partial \tilde{\alpha}_i^{ *}}\right)$. The fifth term is the Lindbladian term describing the dissipation of particles with rate $\gamma$
\begin{align}
\mathcal{T}_{5} =   \sum_i \frac{\gamma_{i}}{2} \left( 2 \hat{b}_i \rho \hat{b}_i^\dagger-\hat{b}_i^\dagger \hat{b}_{i} \rho-\rho \hat{b}_i^\dagger \hat{b}_i \right), \nonumber
\end{align}
which can be mapped to the following equation in the phase-space representation
\begin{align}
    \mathcal{P}_{5} = \int d^{4 M} \vec{\alpha } P \left[-\sum_i \frac{\gamma_{i}}{2}\left( \alpha_i \frac{\partial}{\partial \alpha_i}+\tilde{\alpha}_i^{*} \frac{\partial}{\partial \tilde{\alpha}_i^*}\right)\right] \hat{\Lambda}.\nonumber
\end{align}
The sixth term is also a Lindbladian term describing the particle loss in the source modes. Here, however, the dissipation is directly to the reservoir nodes. Therefore the decay rate needs to be normalised by $\eta = \sum_{i} \left(W_{i}^{\rm in}\right)^{*}$ to ensure particle conservation. The term in the master equation (having a similar form as $\mathcal{T}_{5}$) is given by
\begin{align}
    \mathcal{T}_{6} =  \frac{\gamma_{s}\eta}{2}  f(t)\left(2 \hat{s}_k\rho\hat{s}^{\dagger}-\hat{s}^{\dagger} \hat{s}\rho-\rho \hat{s}^{\dagger} \hat{s}\right).
\end{align}
Analogously, the phase-space integral equation is given by
\begin{align}
    \mathcal{P}_{6} = \int d^{4 M} \vec{\alpha } P \left[-\frac{ \gamma_{s}\eta}{2} \left(s \frac{\partial}{\partial s}+\tilde{s}^* \frac{\partial}{\partial \tilde{s}^*}\right) f(t) \right]\hat{\Lambda}. \nonumber
\end{align}
The final term in the master equation describes how the source modes couple into the reservoir. This coupling is modelled via the cascade coupling formalism, which ensures unidirectional coupling, meaning no back-action from the reservoir to the source mode
\begin{align}
    \mathcal{T}_{7} =  \sum _ { i }\sqrt{\gamma_{s}\gamma_{i}}\sqrt{f(t)} W _ { i } ^ { i n } [ \hat { s }  \rho \hat { b } _ { i } ^ { \dagger } - \hat { b } _ { i } ^ { \dagger } \hat { s} \rho + \hat { b } _ { i } \rho \hat { s} ^ { \dagger } - \rho \hat { s } ^ { \dagger } \hat { b } _ { i } ].\nonumber
\end{align}
The resulting $\mathcal{P}_{7}$ describes how the source modes couple into the reservoir 
\begin{align}
     \mathcal{P}_{7} = \int d^{4 M} \vec{\alpha } P \left[-\sum_{i} \sqrt{\gamma_{s}\gamma_{i}}\sqrt{f(t)} W_i^{i n}\left(s \frac{\partial}{\partial \alpha_i}+\tilde{s}^* \frac{\partial}{\partial \tilde{\alpha}_i^*}\right)\right] \hat{\Lambda}. \nonumber
\end{align}

By integrating the right-hand side equations $\mathcal{P}_i$ within (\ref{eq:integral1}) by parts, one obtains a Fokker-Planck equation under the integrals in (\ref{eq:integral2}). Conveniently, the terms 
for the source modes that go beyond the earlier treatments of \cite{Deuar_2021} are simple, with a zero diffusive term. Since stability was shown to depend almost exclusively on the relationship between occupation, decay and nonlinearity \cite{Deuar_2021}, one can expect similar stability regions to the dissipative Bose-Hubbard model, which turns out to be the case -- see Sec.~\ref{app: stab}.

The drift terms for the source and reservoir are given, respectively, by
\begin{align}
A_{k}&=\left(\begin{array}{cc}-f(t) \frac{\gamma_{s}\eta}{2} s \\  -f(t) \frac{ \gamma_{s}\eta}{2} \tilde{s}^{*}\end{array}\right),\\
A_{i}&=\bigg(\begin{array}{cc}\left[\mathrm{i} \Delta_i-0.5\gamma_i-\mathrm{i} U \alpha_i \tilde{\alpha}_i^*\right] \alpha_i+ \\\left[-\mathrm{i} \Delta_i-0.5\gamma_i  +\mathrm{i} U \tilde{\alpha}_i^* \alpha_i\right] \tilde{\alpha}_i^*+\end{array} \nonumber\\
&\begin{array}{cc}-\mathrm{i} F_i+\mathrm{i}\sum_j J_{ij}\alpha_j - \sqrt{\gamma_{s}\gamma_{i}}\sqrt{f(t)}\,W_i^{\rm in} s \\ \mathrm{i} F_i^*- \mathrm{i}\sum_j J_{ji}^* \tilde{\alpha}_j^* - \sqrt{\gamma_{s}\gamma_{i}}\sqrt{f(t)}\,W_i^{i n} \tilde{s}^{*} \end{array}\bigg).
\end{align}
The diffusion matrix for the reservoir modes comes from the Kerr nonlinearity and is given by\\
\begin{center}
$D_{i i^{\prime}}=\left(\begin{array}{cc}\left(-\mathrm{i} U\right) \alpha_i^2 & 0 \\ 0 & \mathrm{i} U \tilde{\alpha}_i^{* 2}\end{array}\right).$
\end{center}

The diffusion matrix is non-negative, which is necessary to transform the Fokker-Planck Equation to Langevin equations. These equations are effectively the positive-P equations we have been solving in the work and have the form\\
\begin{center}
    $\frac{\partial \vec{v}}{\partial t}=A(\vec{v})+B(\vec{v}) \vec{\xi}(t)$,  \  $B_{i i^{\prime}}=\left(\begin{array}{cc}\sqrt{-\mathrm{i} U} \alpha_i & 0 \\ 0 & \sqrt{\mathrm{i} U} \tilde{\alpha}_i^*\end{array}\right)$,  $BB^{T} = D,$
\end{center}
where $\vec{v}$ is the vector of all fields ($\alpha_j$, $\tilde{\alpha}^{*}_j$, $s$, $\tilde{s}^{*}$), and $\vec{\xi}(t)$ is a vector of uncorrelated Gaussian noises. 

\subsection{Details on quantum state sampling}
The canonical distribution corresponding to any density matrix $\hat{\rho}$ is given constructively by \cite{Drummond_1980}\\
\begin{equation}
P_{\text{can}}(\alpha, \tilde{\alpha}^*) \equiv \frac{1}{4\pi^2} e^{-\tfrac{1}{4}|\alpha - \tilde{\alpha}|^2} \left\langle \tfrac{(\alpha + \tilde{\alpha})}{2} \middle| \hat{\rho} \middle| \tfrac{(\alpha + \tilde{\alpha})}{2} \right\rangle.
\end{equation}
We have used this to generate cat states. For other quantum states portrayed in Figure (\ref{fig:fig8}), the canonical distribution functions, together with a convenient way of sampling these distributions, have been derived in \cite{Olsen_2009}.

For the Schrödinger cat state, we have performed the sampling by first calculating the $P_{\text{cat, can}}\left(\alpha,\tilde{\alpha}^{*}\right)$ distribution on a four-dimensional grid spanned by $\Re{\left(\alpha\right)}$, $\Im{\left(\alpha\right)}$, $\Re{\left(\tilde{\alpha}^{*}\right)}$ and $\Im{\left(\tilde{\alpha}^{*}\right)}$.
The limits of the sampled phase space $\alpha,\tilde{\alpha}$ at large amplitude are taken large enough to make any truncation error negligible. The size of the phase-space thus taken into account can differ depending on the quantum state taken into consideration. Finally, the calculated discrete $P_{cat, can}\left(\alpha,\tilde{\alpha}^{*}\right)$ distribution is sampled and appropriate values of $\alpha$ and $\tilde{\alpha}^{*}$ are chosen.

\begin{figure}[t!]
    \centering
    \includegraphics[width=0.90\linewidth]{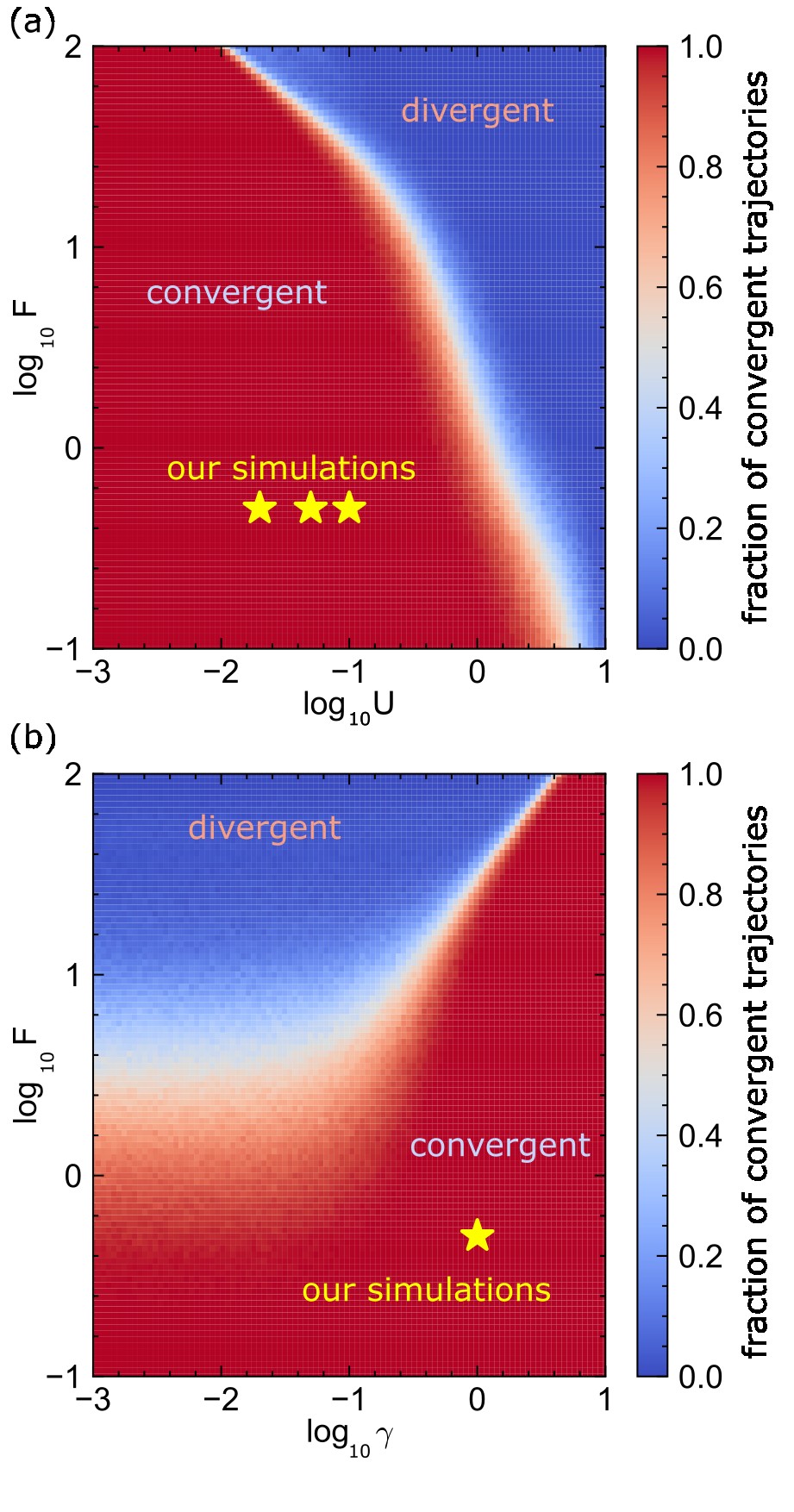}
    \caption{The stability of the positive-$\mathcal{P}$ method for simulating the dynamics of a single reservoir mode for different parameter values. Panel \textbf{a} shows the stability as a function of driving field amplitude $F$ and Kerr nonlinearity $U$. The increase in nonlinearity significantly decreases the stability of the method. The points in the parameter space studied this work have been depicted on the panels with the three points on panel \textbf{a} corresponding to different non-zero Kerr nonlinearities (U = 0.02, 0.05, 0.1). Panel \textbf{b} depicts the stability as a function of driving field amplitude $F$ and dissipation rate $\gamma$. The method becomes unstable for large driving field amplitudes and low losses due to the high average occupation, affecting the noise amplitude.}
    \label{fig:stab-label}
\end{figure}

\subsection{Stability analysis}
\label{app: stab}
Section~\ref{sec:appB} of the Appendix shows that the amplitude of Gaussian noise depends on the Kerr nonlinearity and the amplitude of the corresponding field (e.g. $\alpha$, $\tilde{\alpha}$). A significant contribution of the noise term and low dissipation can contribute to large fluctuations propagating. Therefore, corresponding trajectories may diverge. It is important that we ensure that we are performing simulations in such a region of parameters, where the method is stable. The trajectory number needed for correct convergence may still be large (especially when many trajectories are needed to sample the source state correctly). To qualitatively check this, we have calculated stability diagrams for different cross-sections of the parameter space. We focus on the simplest reservoir consisting of only one mode, without coupling to source states. Consideration of a single mode was found to be fully sufficient for stability analysis of the dissipative Bose-Hubbard model in \cite{Deuar_2021}. The parameters of the reservoir dynamics are the detuning ($\Delta$), Kerr nonlinearity ($U$), pumping amplitude ($F$), and dissipation ($\gamma$). We will be calculating the stability of the method for the following cross-sections: (a) ($F$,$U$) for $\gamma$ = 1 and $\Delta$ = 0, and (b) ($F$,$\gamma$) for $\Delta$ = 0 and U = 0.1. For each point in the parameter space studied, we performed the positive-$\mathcal{P}$ simulation and calculated the fraction of convergent trajectories shown in Figure~\ref{fig:stab-label}. This calculation verified 
that we are working well within the stability regime of the positive-$\mathcal{P}$ method, and this has been consistent with having no diverging trajectories (also in the many-mode runs) when calculating reservoir dynamics described in this work. 
\bibliography{bib}
\end{document}